# Edge Computing Performance Amplification

**Vivek Basavegowda Ramu**[*]

* Independent Researcher, Connecticut, USA.

[vivekgowda.br@gmail.com](mailto:vivekgowda.br@gmail.com)

## Abstract

Edge computing can be defined as an emerging technology that uses cloud computing to leverage edge data centers to process, store, and analyze data close to the source. Traditional cloud computing architectures are not designed for latency-critical applications such as AI (Artificial Intelligence) and IoT (Internet Of Things) because they rely on low data volumes generated by applications running near highly-populated areas. When volume grows beyond 50 miles from the population center, networks experience higher latency and packet loss rates which impacts application performance. Since everyone's life is equipped with more and more IoT devices by the day, decisions should be made at a split second in edge computing. It is really crucial to perform at optimum level, some devices specially the medical wearables deal with patient life and any delay in decision making will result in disaster. Similarly, modern day autonomous self driving vehicles where late decisions can endup in accidents and really there is no room for any errors.

This paper provides a new approach to improve performance of the edge computing by having two identical computing systems in which one system will act as primary and another as reserved or secondary, this system will be available at the local environment of the IoT device and not in cloud. The secondary system will be reserved for mission critical requests and whenever the primary system breaches latency threshold for response only then the request will be re-routed to the secondary system. Both the systems will sync data on background and can also serve as backup computing systems in case of any failure to one of the systems.

Traditional edge computing systems will have a singular computing system on the device and low capability to process data or user requests since it still relies on transferring data to the cloud to compute and make decisions. With multiple high capacity computing systems on the device and with automatic rerouting and balancing of the requests, edge computing performance will be more reliable and it will also ensure high availability, low latency and a highly dependable edge computing architecture. This method is scalable, based on the volume and complexity the architecture can be extended to additional sub systems to add more computing power and to handle additional requests on demand. The proposed method of multiple computing systems at the local environment or device will result in a highly responsive system and provide the much needed support to process data or user requests in a fraction of a second and result in life saving decisions. This solution also opens the door to multiple possibilities like tagging multiple IoT devices to the same computing system, ability to include AI (Artificial Intelligence)/ML (Machine Learning) models and processing locally and to learn from previous decisions to enhance future computing, self troubleshooting and healing process etc which will further advance the existing technology.

* Corresponding Author
  Email Address: [vivekgowda.br@gmail.com](mailto:vivekgowda.br@gmail.com)

## 1.  Introduction

When Akamai introduced CDN`s which stands for content delivery networks in the late 1990s to increase web performance and speed, this is when edge computing first emerged. To cache and prefetch the internet contect, CDN utilized nodes at the edge which are located near to the end users. The information can be changed in a variety of ways by these edge nodes, for as by inserting localized advertising. For video content, CDNs are especially helpful because caching can dramatically reduce bandwidth usage. (Satyanarayanan, M., 2017). The way Edge computing revolutionizes the public cloud is by introducing distributed and decentralized infrastructure, elevating it. By bringing them closer to the target applications, it provides the fundamental foundations of the cloud, including computing, memory, and the network. The latency needed to reach conventional cloud structures is drastically decreased with just one tag to the information center. Edge computing improves the user experience since it reduces the need for a full journey to the cloud. (Janakiram MSV., Jun. 2017)

There is a misunderstanding that edge computing is designed mainly for the IoT. But edge computing is all set to turn into the most desirable structure for running data-driven, sophisticated intelligence based applications. Though edge computing is perfect for IoT solutions, it provides enormous value for the departmental and conventional position of business applications. (Janakiram MSV., Feb. 2017). As the amount of information we are processing continues to expand, we have understood the limitations of cloud technology in some places. Edge computing is introduced to rectify most of these issues by resolving the slowness incurred by computing in the cloud and transporting information to be processed in the data center. It will  reside "on the edge" which is near to where the computing needs to be performed. Because of which, edge computing may best suit to work with time-critical information in remote places which have no reliable connectivity or very limited connectivity to the centralized position. In such places, edge computing in itself will be like mini datacenters.

Even though edge computing has a great reputation in industry for being really fast, the current architecture does have limitations. Sometimes external factors such as computing capacity, software glitch, high-volume processing off request will add high additional overhead on the computing process which will deteriorate the overall response time. In this study an alternative approach for computing which results in better performance is presented. By adding additional system to compute critical requests for the infrastructure we will be able to distribute load and  improve performance drastically, second system will be reserved for only mission critical requests and also serves as disaster recovery system incase of primary system failure.

To provide a better understanding of this topic, the goal of this study is to present the infrastructure enhancement for edge computing which will increase the performance and system stability. This will further give rise to more advancement in IoT devices and computing mechanisms as high volume processing and decision making will be made possible which is the key element for developing new capabilities.

The remaining sections of this paper are organized as follows. Section 2 outlines how cloud and edge computing infrastructure are hosted. Section 3 explains the proposed enhanced edge infrastructure for better performance. Section 4 concluding remarks and summarization of the study. Finally, Section 5 provides limitations and directions for future research.

## 2. Theory/Methodology

A device that is located outside of a data center is referred to as an edge device. Edge computing, because of which, "is a new paradigm in which large processing and storage capabilities are placed at the edge of the Internet, in close proximity to mobile devices or sensors" (the source). Your current mobile phone could be considered an edge gadget. Security cameras may be located across the city and banks. Also regarded as edge devices are self-driving automobiles, thus they don't even need to be small. These gadgets make it simpler and more common to capture real-time and location-based information. In addition, a lot of these gadgets produce actual data like pictures and movies. The enormous volume of data being produced is one of the issues with the proliferation of edge computing devices. Consider a security camera for example, with a 10 Hz frame rate that can transmit data upward of 250MegaBytes each second. An standard airliner with an Internet connection creates data as high as 5 TeraBytes per day. Each day, a self-driving car produces data worth 4TeraBytes and they are only for single devices. Consider the volume of traffic that would result from several of these devices attempting to deliver their unprocessed data to a single server. According to Gartner by year 2025, In contrast to centralized data centers, 75% of data generated by enterprises will be generated elsewhere. The cloud structure is impractical due to issues with bandwidth and latency for devices that need near real-time processing. Energy use is still another issue, a single bit traveling across the Internet uses 500 microjoules of energy, with edge computing this energy usage can be significantly decreased.

As shown in Figure 1, a report from grandviewresearch predicts that the market for edge computing on a global scale was assessed in 2021 for USD 7.43 billion, it is projected to further increase by 38.9% CAGR from the year 2022-2030 (Grand View Research, 2022).

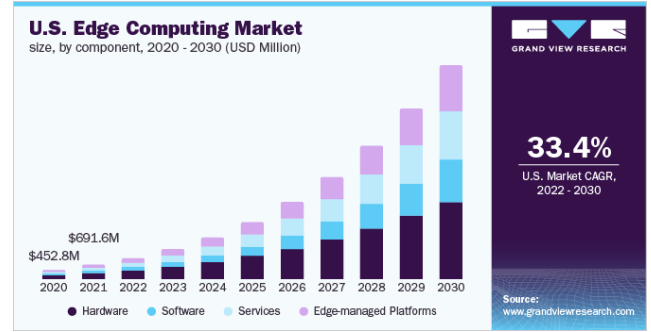

**Figure 1**: U.S. Edge Computing Market Report (Source: Grand View Research, 2022)

### 2.1 How Edge Computing Differs from Cloud Computing

It is really important to understand how edge computing is different from cloud computing before effective performance enhancement is presented. Traditionally software applications always had all infrastructure (servers) on premises and most of the time the infrastructure was under utilized, not easily scalable, and incurred high cost of maintenance. Cloud computing solves this problem by centralizing the infrastructure in a distributed model where one can rent or reserve the required space for operation and only pays based on utilization or for reserved standalone space, this resulted in easy scalability, high uptime and reduction in cost. Due to the longer journey to the server, the likelihood of attacks on the data is greater in cloud computing than in edge computing. The top cloud service providers currently are Amazon Web Services(AWS), Azure from Microsoft and the Google Cloud. Edge computing on the other hand is introduced because of the rise of IoT(Internet of things) devices which requires faster processing, data security and minimum latency, an additional layer of edge node is installed closer to the device which processes requests and controls the device. Table 1 shows the comparison between cloud computing and edge computing (J. Pan and J. McElhannon, 2018).

| Characteristics | Cloud Computing | Edge Computing |
|---|---|---|
| Major applications | Mainstream applications | IoT, VR(Virtual Reality, AR(Augmented Reality), Smart Homes, Smart Vehicles, Smart Devices etc |
| Proximity of services and resources for data processing | Far from end user, in remote location | Closer to user, at the edge |
| Network bandwidth | High, high amount of data transfer is needed | Low, as data processing is done on edge node |
| Availability | A small count of data centers which are large sized | A large count of data centers which are large sized |
| Slowness/ Latency | High, because of the distance | Low, because of the proximity |
| Security | Low | High |
| Scalability | Scalable at data centers in the cloud | Scalable at data centers, at the edge |

**Table 1**: Cloud computing characteristics comparison to edge computing (J. Pan and J. McElhannon, 2018)

## 2.2 Edge Computing Infrastructure

Privacy and security are crucial factors to take into account when using cloud services. Because user privacy must be protected, institutions like hospitals cannot transfer raw data straight to their cloud platforms due to the rising frequency of data breaches. There must be some kind of edge-level preprocessing involved. A simplified 3tier edge computing infrastructure is showcased in Figure 2.

As shown in Figure 3 a typical Edge computing has 3 layers.
- Edge Devices: These are devices used at base level which can be any device like smartphone, smart watch, thermostat, smartplug etc
- Edge Nodes: These are the devices which controls base devices and also collects and processes data like routers, small base stations or data centers which is installed closer to the device
- Cloud: Even though Edge nodes can independently control local devices, when additional communication is required from different parts of the world or to process and analyze data which is not critical, or to update software or security then cloud will be used.

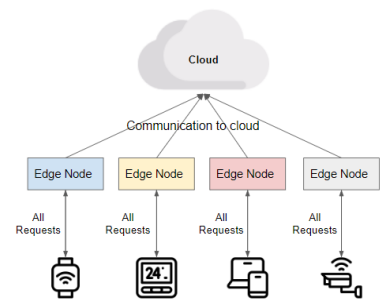

**Figure 2**: Edge Computing Infrastructure

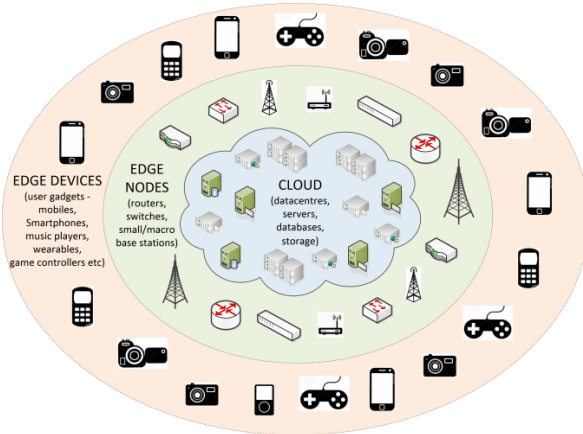

**Figure 3**: Edge Computing Layers (B. Varghese et al., 2016)

As it is evident from the infrastructure setup any disruption on the edge node will result in failure on IoT system and it is not equipped to handle such situation, this clearly indicated there is a need for infrastructure modification which will enable higher availability and performance for edge computing which is the exact goal of this study and explained in detail in next section.

**2.3 Enhanced Edge Computing Infrastructure**

As we have established the limitation with current edge computing infrastructure, the proposal is to have an additional computing system or edge node which will co-exist with the current computing system or edge node. We will consider the existing edge node to be the primary node and newly added node to be secondary node. Whenever the primary node fails, is not available or has high latency, the secondary node will be active and will process mission critical requests to ensure the IoT device can continue the operation.

In Figure 4 it is showcased that each of the IoT device will talk only to the primary node for regular requests and processing and critical requests are either directly sent to secondary node or when the primary node fails or is non-responsive. Both the edge nodes will be in-sync with near zero latency if there is any need of data sharing for request processing. This can be achieved by tagging the request as critical so that when routed it can appropriately be sent to the primary or secondary node.

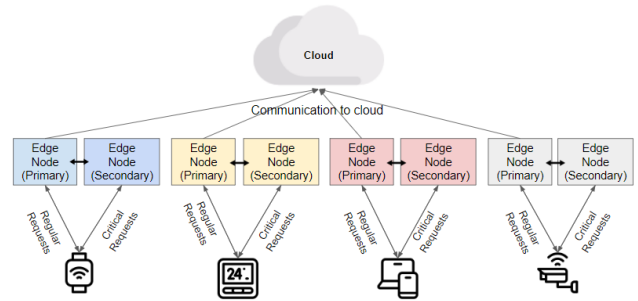

**Figure 4:** Enhanced Edge Computing Infrastructure

## 3. Results

To elaborate on the proposed enhanced infrastructure advantage, we will consider 3 different scenarios. Response time numbers used are not actual tested numbers and it is for the purpose of better demonstration of the proposed method. In general for interactive applications, reaction delays of less than 150 ms are typically regarded as acceptable; however, fast-paced engagements cannot tolerate delays of more than 70 ms (Premsankar, G. et al., 2018).

**3.1 Scenario 1 : Primary Node is Active**

This is the happy path, as shown in Figure 5 when the primary node is actively processing all requests and does not have any latency impact. In this scenario, the primary node will continue to process requests and the secondary node will be in stand-by mode. Only when the latency threshold is breached by the primary node the request is killed and resent to the secondary edge node for processing. Latency threshold is a variable value and can be set based on the criticality of the process, as a ballpark number any request which does not receive response beyond 100ms can be considered to be redirected.

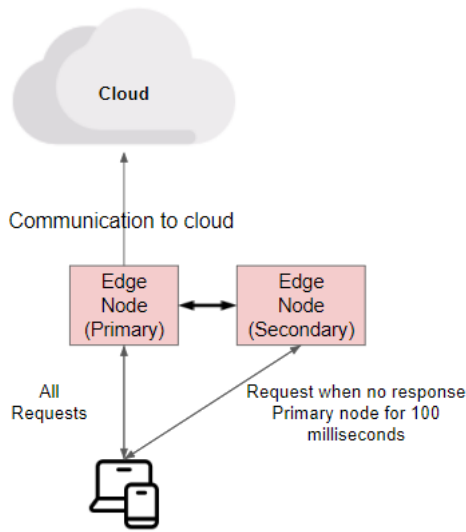

Figure 5: Scenario 1 - Primary Node is Active

### 3.2 Scenario 2 : Primary Node and Secondary Node is Active

As shown in Figure 6, in this scenario we will balance the requests for enhanced performance by routing critical requests only to secondary node and all other majority of non-critical requests will be sent to primary edge node. By doing this we will ensure the mission-critical requests are always served without any latency which will drive customer satisfaction and prolonged service to the devices. This is a better approach compared to Scenario-1 because we will be utilizing both the nodes for optimal performance.

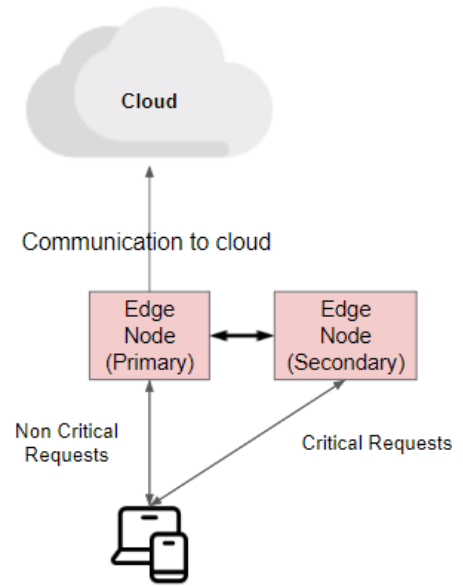

Figure 6: Scenario 2 - Primary Node and Secondary Node is Active

### 3.2 Scenario 3 : Primary Node is Inactive

When Primary node fails to process any requests, this could be due to many reasons like memory issues, software glitches, security or software updates processing etc. Then the secondary edge node will act as a disaster recovery system, meaning it will take the burden of processing all the requests to serve the IoT device to have no impact on the service. Here the secondary node will in turn become the primary node till the actual primary node is available to process requests again, this scenario is shown in Figure 7.

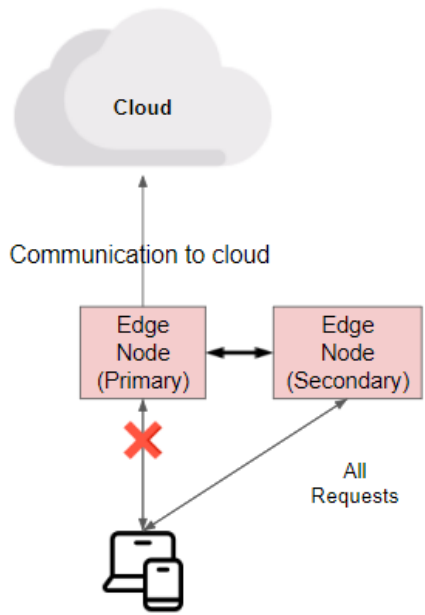

**Figure 7**: Scenario 3 - Primary Node is Inactive

In Table 2, the different scenarios are listed mapping to the status of primary and secondary nodes along with the type of request which is expected to be processed by the secondary node. This gives a comprehensive view of the edge nodes' expected activity when proposed infrastructure is implemented.

| Scenario | Primary Node Status | Secondary Node Status | Secondary Node Request Process Type |
|---|---|---|---|
| Scenario 1 : Primary Node is Active | Active | Inactive | Critical requests on demand only |
| Scenario 2 : Primary Node and Secondary Node is Active | Active | Active | Critical requests only |
| Scenario 3 : Primary Node is Inactive | Inactive | Active | All requests |

**Table 2**: Edge Node status and Secondary node request process type

## 4. Conclusion

Now that we have IoT, we have reached the post-Cloud age, when there will be a lot of edge apps deployed and a lot of high-quality data generated by devices that are part of our daily lives (Shi, W. et al., 2016). IoT will be the most crucial component of big data production because it is the supply of enormous amounts of data. IoT must therefore upload vast amounts of data to edge or cloud storage. Of course, the quick upload time is one advantage of using edge-based storage (Yu, W. et al., 2017). Incorporating edge computing with 5G technology has the potential to be extremely beneficial for the healthcare industry, as it will promote advancements in remote operations and diagnostics as well as the tracking of patient's vital signs and statistics. Doctors can operate surgical instruments remotely from a location where they feel secure and comfortable in order to save lives (Hassan, N. et al., 2019). On the other hand, edge systems do create a lot of data which may not be required to be stored or transferred to cloud, consider medical devices which continuously monitor patients vitals, most of the time the stable measurement can be ignored and only when the readings are at critical level are most important.

With the way edge computing is booming it is really crucial to design effective infrastructure which can handle any circumstances, work under network limitations, intelligently routes requests on-demand, minimize the dependency on large cloud computing systems and can be highly reliable, with the current infrastructure setup for most of the IoT devices it is nearly impossible to completely rely on one edge node or computing system to serve all requests effectively and in majority of the cases the increased latency will have significant impact. In this study we identified the limitations in current edge computing infrastructure and have successfully presented an enhanced infrastructure which will have additional edge node to compute critical requests or serve as backup computing system incase of any disaster to ensure no disruption to the way IoT devices operate. If effectively implemented this enhanced infrastructure will showcase a lot of benefits in terms of fast response, high uptime and reliability. This will also give rise to future sets of new innovations in the edge computing ecosystem by the invention of new devices and highly processing units which can make concepts like metaverse a reality much sooner.

## 5. Discussion

Edge computing is considered to be the key transformer for a number of developing technologies, such as augmented reality, 5G, vehicle-to-vehicle communications and the Internet of Things (IoT) by connecting the end users to cloud computing resources and services (Khan, W.Z. et al., 2019). Eventhough enhanced infrastructure will have higher edge computing capacity there are some limitations related to additional cost involved to setup secondary edge node, data sync between both primary and secondary edge nodes at near zero latency can be challenging, actual setup and testing is required to solidify the proposed solution. One other aspect of edge computing which cannot be ignored is the physical maintenance. Unlike cloud, the edge systems which do computing close to the devices should be maintained physically overtime. Security of physical devices is also another area of edge computing which is vulnerable. In General, there is an increase in data breach incidents and more hackers are coming into play day-by-day, be it On-prem, Cloud, Hybrid or Edge Infrastructure. In one such incident, recently Uber company data was hacked (Vivek Basavegowda Ramu, 2022 ).

An interesting future research direction would be to test secondary and additional edge nodes. Such a study could consider, for instance, setting up additional edge nodes to find the optimal cost to performance ratio, connect multiple devices to edge node clusters etc. We hope that this paper will encourage further research in this direction.